\documentclass[conference]{IEEEtran}
\IEEEoverridecommandlockouts
\usepackage{cite}
\usepackage{amsmath,amssymb,amsfonts}
\usepackage{algorithmic}
\usepackage{graphicx}
\usepackage{textcomp}
\usepackage{xcolor}
\usepackage{tablefootnote}

\usepackage[misc]{ifsym}
\usepackage{graphicx}
\usepackage{subfig}
\usepackage{adjustbox}
\usepackage{cite}
\usepackage{amsmath,amssymb,amsfonts}
\usepackage{algorithmic}
\usepackage{graphicx}
\usepackage{textcomp}
\usepackage{xcolor}
\def\BibTeX{{\rm B\kern-.05em{\sc i\kern-.025em b}\kern-.08em
    T\kern-.1667em\lower.7ex\hbox{E}\kern-.125emX}}
\begin{document}

\title{Split Federated Learning: Speed up Model Training in Resource-Limited Wireless Networks}
\vspace{-0.1cm}
\author{\IEEEauthorblockN{Songge Zhang$^{*,\star}$, 
     Wen Wu$^{\star\textrm{,\;\Letter}}$, 
      Penghui Hu$^\dagger$,
       Shaofeng Li$^\star$, and Ning  Zhang$^\ddagger$}\\
    \vspace{-0.3cm}
        \IEEEauthorblockA{
    Peking University Shenzhen Graduate School, Peking University, Shenzhen, China$^*$\\
      Frontier Research Center, Peng Cheng Laboratory, Shenzhen, China$^\star$\\
      School of Electronic Science and Engineering, Nanjing University, Nanjing, China$^\dagger$\\
     Department of Electrical and Computer Engineering, University of Windsor, Windsor, Canada$^\ddagger$\\
     Email: \mbox{zhangsongge@stu.pku.edu.cn}, \mbox{\{wuw02, lishf\}@pcl.ac.cn}, \mbox{huph197@smail.nju.edu.cn}, and \mbox{ning.zhang@uwindsor.ca}
    }
\vspace{-0.5cm}
}

\maketitle
\begingroup\renewcommand\thefootnote{${\textrm{\Letter}}$}
\footnotetext{Wen Wu (wuw02@pcl.ac.cn) is the corresponding author of this paper.}
\endgroup

\begin{abstract}
In this paper, we propose a novel distributed learning scheme, named group-based split federated learning (GSFL), to speed up artificial intelligence (AI) model training. Specifically, the GSFL operates in a \emph{split-then-federated} manner, which consists of three steps: 1) Model distribution, in which the access point (AP) splits the AI models and distributes the client-side models to clients; 2) Model training, in which each client executes forward propagation and transmit the smashed data to the edge server. The edge server executes forward and backward propagation and then returns the gradient to the clients for updating local client-side models; and 3) Model aggregation, in which edge servers aggregate the server-side and client-side models. Simulation results show that the GSFL outperforms vanilla split learning and federated learning schemes in terms of overall training latency while achieving satisfactory accuracy.
\end{abstract}

\section{Introduction}
Distributed learning paradigms, e.g., federated learning (FL) and split learning (SL) \cite{ref1}, have been widely applied in the fields of healthcare, finance, and autonomous driving. FL enables parallel training of a shared artificial intelligence (AI) learning model on the local dataset of participating clients and only delivers the model parameters without sharing raw data. FL suffers from significant communication overhead issues in resource-limited wireless networks due to uploading large data-size AI models. SL splits an AI model into two parts at a cut layer, i.e., a client-side model on the client and a server-side model on the edge server. With the participation of multiple clients in SL, long training latency becomes a significant issue \cite{ref3}. Thus, there is a need for designing a novel scheme that can facilitate parallel model training to reduce training latency.

A hybrid federated split learning scheme can reduce communication overhead and computation overload on clients \cite{ref6}. However, the simple combination scheme requires equipping each client with a server-side model. When there are many clients, the number of server-side models is large, consuming prohibitive storage resources. Therefore, a new resource-efficient distributed learning scheme is required.

In this paper, we propose a group-based split federated learning scheme (GSFL) to reduce training delay. The advantage of a group-based strategy is that it can facilitate parallel model training across different clients. This strategy circumvents the sequential training process of all clients, thereby accelerating the overall training process. Specifically, the proposed scheme employs a split-then-federated manner to train clients. In the inter-group, clients sequentially interact with the server to complete split training. In the intra-groups, the groups can share the same server-side model for parallel training to improve training efficiency. Thus, the proposed GSFL scheme can speed up the training process compared with conventional SL and FL. Simulation results show that the proposed scheme can reduce training latency, as well as converge to satisfactory accuracy.

\begin{figure}[!t]
\centering
\includegraphics[width=0.39\textwidth]{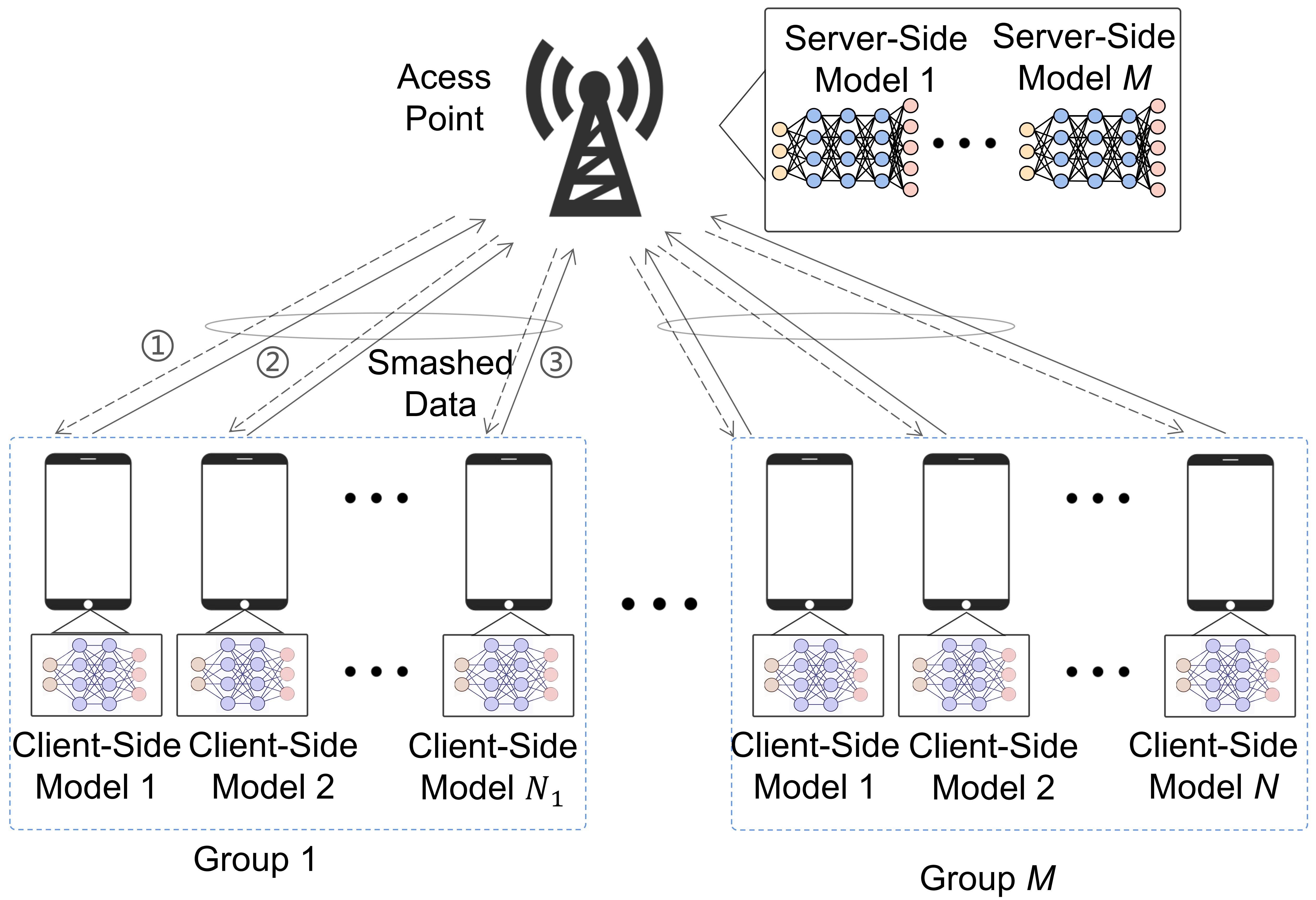}

\caption{Proposed GSFL framework.}
\vspace{-0.5cm}
\label{fig:example}
\end{figure}

\section{Group-Based Split Federated Learning Scheme }

As shown in Fig. \ref{fig:example}, we consider a generic wireless network scenario, comprising one access point (AP) and $N$ clients, i.e., mobile devices. To reduce the communication overhead and computational overload, we partition the clients into $M$ groups. Clients within each group do not share their local data but collaboratively train a local AI model. At the AP, an edge server is deployed, featuring abundant computation and storage resources. Concurrently, the edge server hosts multiple identical server-side models.

The proposed scheme mainly consists of three steps: model distribution, model training, and model aggregation. The main purpose of model distribution is to distribute the client-side model to clients. Model training involves model execution, model updating, and model sharing, which is to complete the split learning process. Model aggregation is performed at the AP, where the well-trained models are aggregated into the whole model.

\subsection{Step 1: Model Distribution}
In the model distribution process, the AP adopts a partitioning strategy to partition the original AI training model into distinct segments. This partitioning allows the AP to obtain the initial server-side models and client-side models. The client-side models can be distributed to the first-trained clients in each group using wireless communication links. 

\subsection{Step 2: Model Training in Each Group}
\textbf{1. Model execution:~}During the model execution phase of the training process, data sampling, client-side model forward propagation, and server-side model forward propagation are necessary. Specifically, in each training epoch for the $n$-indexed client of the $m$-indexed group, a mini-batch of data samples are sampled from the local data set. Then, the client executes the forward propagation process of model training, inputs the sampled data into the local model, obtains the inference output \textquotedblleft smashed data\textquotedblright of the segmentation layer, and transmits it to the AP. Upon receiving the smashed data from the client, the AP inputs it into the server-side model along with prediction results, thereby completing the forward propagation process of the server-side model.

\textbf{2. Model updating:~} In the process of updating the model, updates are required for both the server-side and the client-side models. Specifically, the AP can compute the loss function and its gradient based on the prediction results obtained through forward propagation. The update of the server-side model parameters can be accomplished through methods such as stochastic gradient descent, which completes the backward propagation of the server-side model training. Then, the loss function gradient of the smashed data is transmitted to the corresponding client in the group. Upon receiving the gradient, each client-side model is updated locally, which completes the backward propagation of the client-side model.

\textbf{3. Model sharing:~}In the process of model sharing, the well-trained client-side model is delivered to the next client through the AP. After all clients in each group have completed local training, the last-trained client sends the client-side model to the AP. 

\subsection{Step 3: Model Aggregation among Groups}

After all groups have completed the model training process, AP aggregates all the server-side models and client-side models into a new one respectively. As such, one training round of the AI model is accomplished. Model aggregation can be conducted through FedAVG. The aggregated model is used in the next training round until satisfactory accuracy is reached.

\begin{figure}[!t]
\centering
\begin{adjustbox}{max width=\textwidth}
\begin{tabular}{@{}cc@{}}
\subfloat[Accuracy]{\includegraphics[width=0.5\linewidth]{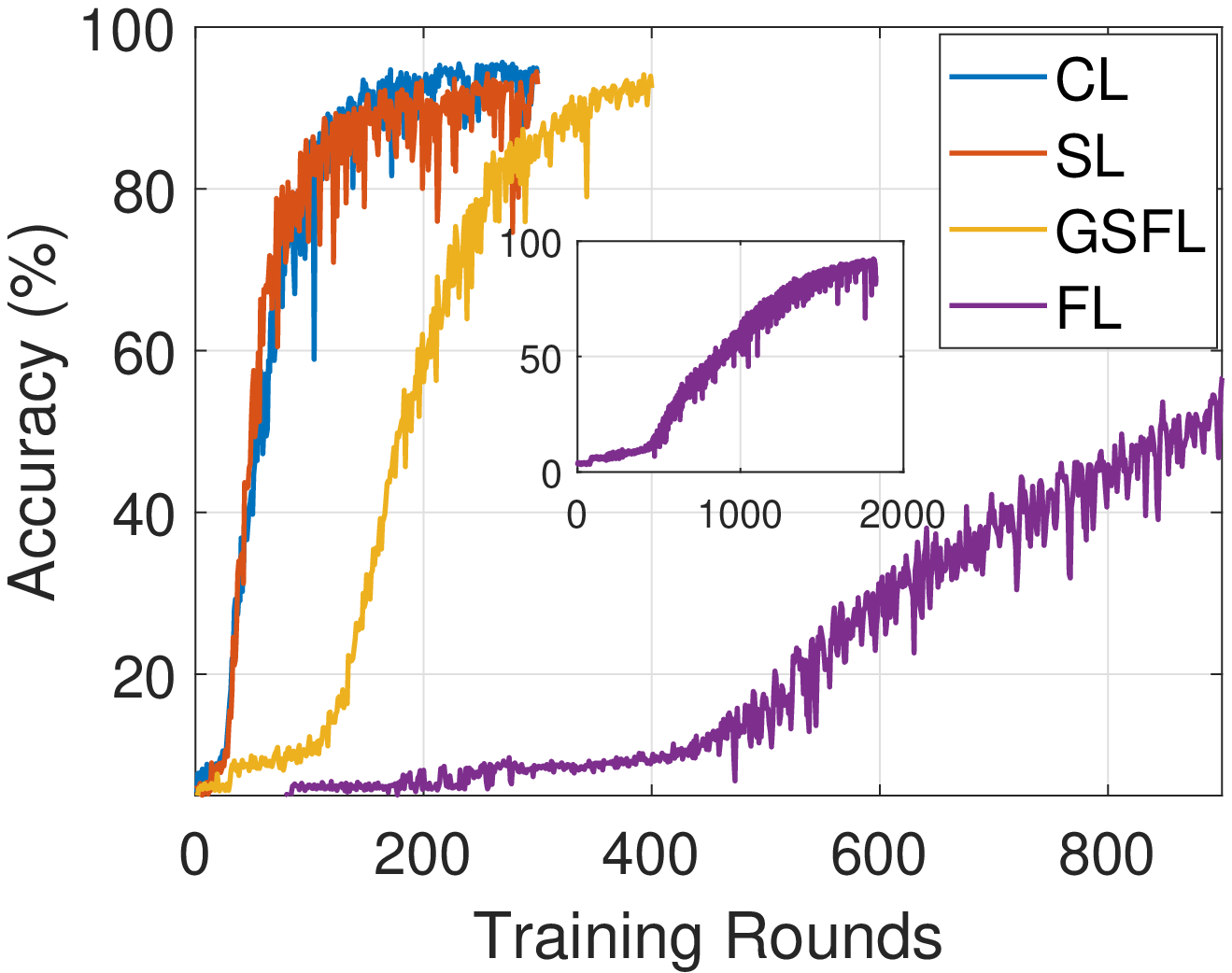}%
\label{case1}}\hspace{-4mm} &
\subfloat[Delay]{\includegraphics[width=0.5\linewidth]{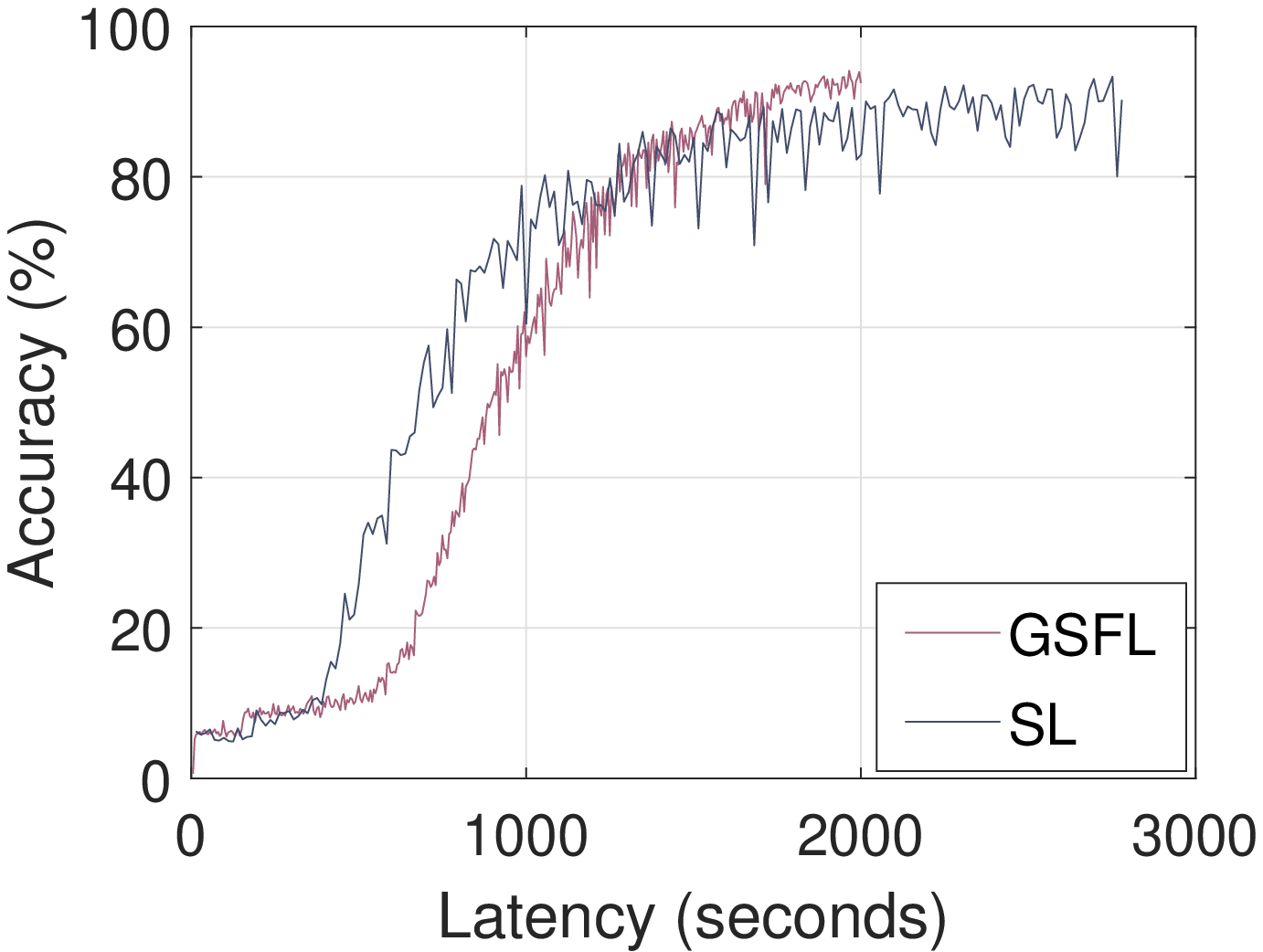}%
\label{case2}}\\[-2mm]
\end{tabular}
\end{adjustbox}
\caption{Performance comparison with respect to training accuracy and training latency on the GTSRB dataset.}
\vspace{-0.3cm}
\label{fig_sim}
\end{figure}

\section{Initial Simulation Results}

We conduct simulations to evaluate the performance of the proposed GSFL scheme on the GTSRB dataset \cite{ref5}. The considered network consists of 30 clients, which are divided into 6 groups. We compare the proposed GSFL scheme with three benchmark schemes: centralized learning (CL), vanilla SL, and FL. Fig. \ref{fig_sim}(a) presents the comparison results in terms of accuracy, showing the accuracy performance with respect to the number of training rounds. The simulation results demonstrate that the proposed GSFL scheme achieves an accuracy level comparable to that of the SL scheme and CL scheme, indicating its effectiveness in model training. The proposed GSFL scheme requires more training rounds to converge compared to benchmark two schemes due to the model aggregation among groups. In addition, the proposed GSFL scheme exhibits significant advantages over FL, with nearly $500\%$ improvement in convergence speed, indicating the superiority of the proposed scheme. To evaluate the delay performance, we compare the proposed GSFL scheme and the vanilla SL scheme in Fig. \ref{fig_sim}(b). It can be seen that the GSFL achieves faster convergence and shorter delay compared to the SL scheme, indicating its superiority in optimizing the global model with reduced training delay. Specifically, the proposed GSFL scheme reduces the delay by about $31.45\%$, demonstrating superior performance in terms of speeding up model training. 

\section{Conclusion and Future Work}

In this paper, we have validated that our proposed scheme can greatly reduce training latency while preserving high model accuracy. For the future work, we will study the impact of the cut layer selection and client grouping on the system performance. Meanwhile, rationally allocating communication bandwidth and computing resource is crucial for enhancing system performance, and hence we will design an effective resource allocation scheme for the proposed scheme.

\section*{ACKNOWLEDGEMENT}
This work was supported in part by the Peng Cheng Laboratory Major Key Project under Grant PCL2021A09-B2 and by the Natural Science Foundation of China under Grant 6220012314.

\bibliographystyle{IEEEtran}

\bibliography{ref.bib}

\end{document}